\documentclass[authoryear]{elsarticle}

\usepackage{graphicx}
\usepackage{amsmath,amsfonts,amssymb,dsfont}
\usepackage[english]{babel}
\usepackage{color}
\usepackage{colortbl,rotating}
\usepackage{natbib}

\newcommand{\erre}{{\ttfamily{{\bf R}\,}}}
\newcommand{\bX}{\mathbf{X}}
\newcommand{\bxi}{\text{\mathversion{bold}$\xi$\mathversion{normal}}}
\newcommand{\bomega}{\text{\mathversion{bold}$\omega$\mathversion{normal}}}
\newcommand{\bOmega}{\text{\mathversion{bold}$\Omega$\mathversion{normal}}}
\newcommand{\balpha}{\text{\mathversion{bold}$\alpha$\mathversion{normal}}}
\newcommand{\bx}{\mathbf{x}}
\newcommand{\bPhi}{\text{\mathversion{bold}$\Phi$\mathversion{normal}}}
\newcommand{\reals}{\mathbb{R}}
\newcommand{\bSigma}{\text{\mathversion{bold}$\Sigma$\mathversion{normal}}}
\newcommand{\by}{\mathbf{y}}
\newcommand{\bY}{\mathbf{Y}}
\newcommand{\bS}{\mathbf{S}}
\newcommand{\bB}{\mathbf{B}}
\newcommand{\bR}{\mathbf{R}}
\newcommand{\bV}{\mathbf{V}}
\newcommand{\bC}{\mathbf{C}}
\newcommand{\bM}{\mathbf{M}}

\newcommand{\bz}{\mathbf{z}}
\newcommand{\bDelta}{\text{\mathversion{bold}$\Delta$\mathversion{normal}}}
\newcommand{\bdelta}{\text{\mathversion{bold}$\delta$\mathversion{normal}}}
\newcommand{\bU}{\mathbf{U}}
\newcommand{\bG}{\mathbf{G}}
\newcommand{\bpsi}{\text{\mathversion{bold}$\psi$\mathversion{normal}}}
\newcommand{\bte}{\text{\mathversion{bold}$\theta$\mathversion{normal}}}
\newcommand{\bLambda}{\text{\mathversion{bold}$\Lambda$\mathversion{normal}}}
\newcommand{\btheta}{\text{\mathversion{bold}$\theta$\mathversion{normal}}}
\newcommand{\bTheta}{\text{\mathversion{bold}$\Theta$\mathversion{normal}}}
\newcommand{\bm}{\mathbf{m}}
\newcommand{\expe}[1]{\operatorname{\rm I\hspace{-.25em}E}\left(#1\right)}
\newcommand{\iid}{\stackrel{\texttt{iid}}{\sim}}
\newcommand{\diag}[1]{\operatorname{\rm diag}\left(#1\right)}
\newcommand{\mez}{\frac 12}
\newcommand{\bzero}{\text{\mathversion{bold}$0$\mathversion{normal}}}
\newcommand{\beq}{\begin{equation}}
\newcommand{\enq}{\end{equation}}
\newcommand{\bei}{\begin{itemize}}
\newcommand{\eni}{\end{itemize}}

\newcommand\independent{\protect\mathpalette{\protect\independenT}{\perp}}
\def\independenT#1#2{\mathrel{\rlap{$#1#2$}\mkern2mu{#1#2}}}
\def\boldfacefake #1{%
        \hbox{%
                \mathsurround=0pt
                \hbox to 0.25pt{$#1$\hss}%
                \hbox to 0.25pt{$#1$\hss}%
                \hbox {$#1$}%
        }%
}

\newtheorem{propo}{Proposition}

\begin{document}

\begin{frontmatter}
 \journal{Computational Statistics \& Data Analysis}
 \title{Bayesian inference for the multivariate skew-normal model: a Population Monte\,Carlo approach}
 \author[BL]{Brunero Liseo}
 \address[BL]{MEMOTEF, Sapienza Universit\`a di Roma}
 \author[AP]{Antonio Parisi\corref{cor1}}
 \ead{antonio.parisi@uniroma2.it}
 \cortext[cor1]{Via Columbia 2 - 00133 Roma - Italy, 
 Tel: +39 06 72 59 59 14;\, Fax: +39 06 20 40 219}
 \address[AP]{Dipartimento DEF, Universit\`a degli Studi di Roma ``Tor Vergata''
 }

\begin{abstract}
Frequentist and likelihood methods of inference based on the 
multivariate skew-normal model
encounter several technical 
difficulties with
this  model. 
In spite of the popularity of this class of densities, there are no broadly satisfactory
solutions for estimation and testing problems.  A general
population Monte Carlo algorithm is proposed which: 1) exploits the latent structure
stochastic
representation of skew-normal random variables to provide
a full Bayesian analysis of the model and 2) accounts for the presence of
constraints in the parameter space. The proposed approach can be defined as weakly
informative, since the prior distribution approximates the actual reference prior for the
shape parameter vector.
Results are compared with the existing classical solutions and the practical
implementation of the algorithm is illustrated via a simulation study and a real data
example.
A generalization to the matrix variate regression model with skew-normal error is also
presented.
\end{abstract}
\begin{keyword}
 Bayes factor \sep Matrix variate regression \sep Objective
Bayes inference \sep Population Monte Carlo \sep Reference prior \sep Skewness
\end{keyword}
%{\bf AMS classification:} 62F15.
\end{frontmatter}

\section{Introduction}\label{intro}
The skew-normal (SN hereafter) class of densities has independently and recurrently appeared
in statistical literature: see for example 
\cite{rob:66} and \cite{aoh:76}; it was named by \citet{azz:85} and further generalised to
the multivariate case by \cite{adv:96} and \cite{ac:99}. The
appearance of the multivariate version is to be considered the starting point of a
dramatically prolific line of
research, both from a methodological and an applied perspective. 
Comprehensive accounts of the huge production of papers and applications related to
the SN model and its ramifications can be found,
for example, in the book edited by \cite{gen:04}, or in the review paper by
\cite{azz:05}. 
The popularity of this class of distributions stems mainly from
its ability to capture and explicitly model mild departures from symmetry,
without losing mathematical tractability, which can be particularly useful in real data
applications.
Another reason for the popularity of the SN class is because it
naturally arises in real data analysis under
special mechanisms of data collection, such as hidden truncation or selective
reporting: see \cite{barry}.
A deeper analysis of the literature, however, reveals that most of the
existing results are 
restricted to the distributional theory of skew-normal and, more generally,
skew-elliptical distributions. 
On the other hand, the theory of inference is still
problematic even in the
scalar case \citep{ac:03}. These problems were anticipated in \citet{azz:85} and
\citet{lise:90}, and are basically due to a number of anomalies of the likelihood
function:
for instance, under the scalar stndard skew-normal model, there is a positive sampling probability
that the maximum likelihood estimator will produce infinite values; specifically, this
phenomenon occurs when all the data points have the same sign. 
These difficulties tend to be more challenging in the multivariate set-up where, in
addition, ``problematic'' situations are not so easy to detect. 
Even ignoring these pathological cases, the likelihood surface arising from an i.i.d.
sample of skew-normal random variables is often non regular and maximum likelihood 
estimates (MLE, hereafter) 
tend to be unstable.

In this paper
we describe a full Bayesian analysis of the multivariate SN
model. In particular we propose:
\bei
\item to use objective priors, in order to correct the odd behavior of the likelihood
function without introducing external information;
\item to exploit the latent structure of the SN model in order to tailor a specific
version of a Population MonteCarlo (PMC, hereafter) algorithm,
and to produce valid posterior inferences, in terms of estimation and testing.
\eni

The paper is organized as follows: Section \ref{mot} introduces the multivariate SN
model and presents a few examples that motivates the proposal of the paper. Section \ref{22}
introduces an augmented
likelihood function 
which exploits the intrinsic latent structure of the skew-normal model. In Section
\ref{Prior distributions} 
we discuss the choice of prior distributions; in Section \ref{Population Monte Carlo
algorithm} 
we describe a PMC algorithm 
with proposal densities based on the full conditional distributions of the parameters
\citep{MR2236856};
in Section \ref{4} we discuss the testing and model selection problems, where a comparison
between the nested normal and the skew-normal model may be of interest. 
Section \ref{5} generalises the approach to the matrix variate regression model,
which is useful when a set of covariates is available. We also discuss some technical
and practical issues related to the algorithm. Finally, Section \ref{Examples} deals
with some numeric comparisons with other existing methods and the analysis of a financial
data set.

\section{Motivations}
\label{mot}

A random vector $\bX$ is said to have a $p$-dimensional standard SN distribution, with
correlation 
matrix $\bOmega$ and shape parameter $\balpha$ when its density function is
\beq
\label{uno}
f\left( \bx; \bOmega ,\balpha \right) =2\varphi_{p}\left( 
\bx;\bOmega \right) \bPhi_1 \left[ \balpha^\prime
\bx \right], \qquad \bx, \balpha \in \reals^{p},
\enq 
with $\varphi_p({\bf w}, {\bf A})$ denoting the density of a $p$-dimensional normal random vector
with standard marginals and covariance matrix ${\bf A}$, evaluated at ${\bf w}\in \reals^{p}$, and 
$\bPhi_1({\bf w})$ is the cumulative distribution function of a standard scalar normal random variable. 
Note that $\bOmega$ is a correlation matrix, although it is not the correlation matrix for
the components of $\bX$; it even appears in the standard version of the SN model. It is
easy to generalise the model with the inclusion of location and scale parameters.
Let $\bxi$ be a $p$-dimensional vector of real numbers and
$${\bf \bomega} = \diag{\omega_1, \dots, \omega_p}
$$
be a diagonal matrix with the marginal scale parameters, so that $\bSigma=
\bomega \Omega \bomega$ represents the scale matrix;
then $\bY= \bxi + \bomega \bX $ has a $p$-dimensional SN distribution ($SN_p(\bSigma,\bxi,
\balpha)$, hereafter) 
with density
$$
f(\by; \bxi, \bSigma, \balpha) = 
2 \varphi_p (\bx-\bxi; \bSigma) \bPhi_1\left[\balpha^\prime \bomega^{-1}(\bx-\bxi)\right ].$$
In this parameterization, each component of the shape parameter $\balpha$ can take any
real value. 
An alternative parameterization \citep{ac:99}, defined in terms of $\bdelta$,
exists, namely
\beq
\label{2}
\balpha =(1-\bdelta^\prime\bOmega^{-1}\bdelta)^{-\mez}\bOmega^{-1}\bdelta,
\enq
or equivalently,
\beq
\label{2delta}
\bdelta =(1+\balpha^\prime\bOmega\balpha)^{-\mez}\bOmega \balpha.
\enq
Notice that, although each component $\delta_j$ takes values in $[-1,1]$, the entire
vector $\bdelta$ belongs to an ellipsoidal subset of $[-1,1]^p$ whose shape is
regulated by $\bOmega$. Although 
this problem is crucial in any simulation based Bayesian approach for inference,
it seems to have been neglected in the literature; we will return to this issue below.
Another possible parameterization, which is particularly useful for likelihood-based
inference, has been proposed in \cite{ava:08}. 

Consider now the simplest inferential situation, where one observes an i.i.d.
sample 
$\by=(y_1, \dots, y_n)$ 
of $n$ observations from an $SN_p(\bSigma,\bxi, \balpha)$ population.
The likelihood function is then 
\begin{eqnarray}
\label{likel}
{L(\bSigma, \bxi, \balpha; \by)}&\propto&
\mid \bSigma \mid^{-\frac n2} 
\exp
\left \{ 
-\frac 12 \sum_{i=1}^n 
\left [  ({\by_i} - \bxi )^\prime \bSigma^{-1} ({\by_i} - \bxi ) \right ] 
\right \} \notag \\ 
&\times & 
\prod_{i=1}^n \bPhi_1\left(\balpha^\prime \bomega^{-1} (\by_i-\bxi)\right ).\notag 
\end{eqnarray}
This likelihood function is quite difficult to manage \citep{ac:99}: 
there are no closed form expressions for the maximum likelihood estimator and, as
anticipated, the MLE of $\balpha$ can be infinite even in very simple settings. 
Consider, for example, the case $p=2$, when all the parameters except $\balpha$
are known: suppose we observe the following bivariate random sample of size $10$; the
first
(second) 
row indicates $X_1$ ($X_2$) values:
{\small
$$
\begin{bmatrix}
 -0.272  & 0.340 & 0.498 & 1.511 & -0.134 & 0.170 &-0.169 & 0.484 & -1.042 & 0.945 \\
 1.421   & 0.668 & 1.610 &-0.610 & 0.577 & -0.168 &  2.222 & -0.606& 1.789 & 0.361 \\
\end{bmatrix}.
$$
}
\begin{figure}[ht]
\centerline{\includegraphics[angle=-90,width=0.7\textwidth]{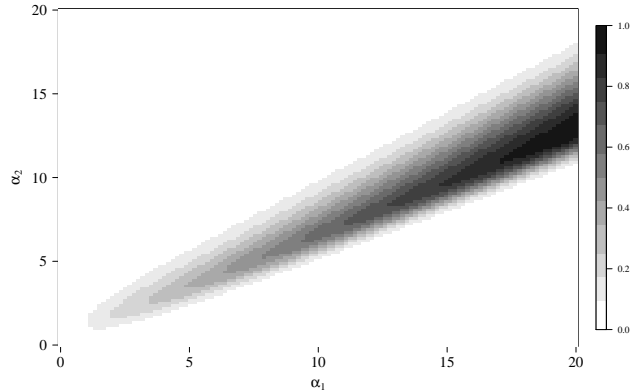}}
 \label{fig:con_plot}
 \caption{An example of an ill-behaved likelihood function.}
\end{figure}

\noindent Figure \ref{fig:con_plot} depicts the contour plot of the likelihood function 
for $\balpha=(\alpha_1, \alpha_2)$; it is clear that the MLE of the vector $\balpha$ is
infinite: the \erre function \texttt{msn.mle} in the suite \texttt{sn}
provides the 
estimates
$$
(\hat{\alpha}_1, \hat{\alpha}_2 ) = (24776144, 19911143).
$$
The unsatisfactory behaviour of
the maximum likelihood method is not immediate clear from the sample values. 
Table 3 in \cite{eling} shows an even more dramatic example with real data in the
context of the skew-$t$ model.

\begin{figure}[ht]
 \centerline{\includegraphics[width=\textwidth,width=0.7\textwidth]{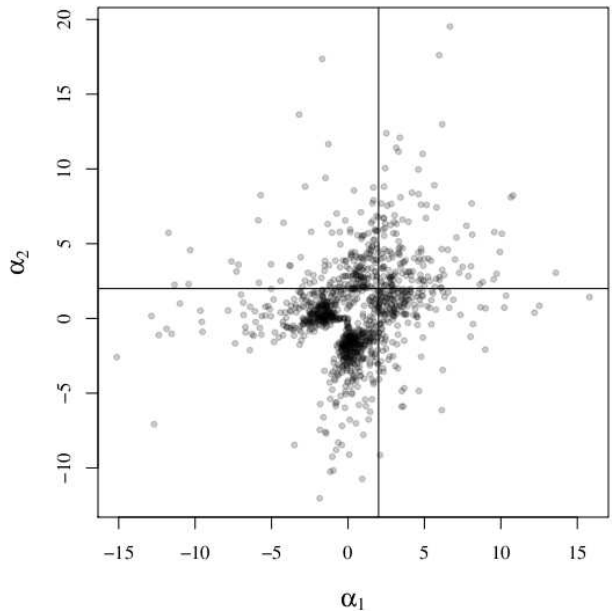}} 
 \caption{Sampling distribution of the (finite-valued) maximum likelihood estimates 
 of $\balpha$. Solid lines indicate the true value.}
 \label{fig:stime}
\end{figure}

To emphasize this point, we have generated 2000 samples of size 30 from a $SN_2$ density
with $\bxi=(0,0)$, $\Sigma=I_2$ and $\balpha=(2,2)$.
Point estimates of the shape vector have been obtained, based on the \erre suite {\tt sn},
which can be considered as a benchmark in this context.
Out of 2000 samples, about $38\%$ resulted in an infinite estimate for
$\balpha$;
Figure \ref{fig:stime} shows the subset of the finite point estimates for $\balpha$. Of
course, this admittedly unsatisfactory behaviour tends to be even worse for smaller
sample sizes and/or for larger values of $\balpha$. 
While in the scalar case the set of samples producing infinite ML estimates of
$\alpha$
 (or $\mid\delta\mid=1$) can be exactly characterized \citep{lisl:05}, the detection of
such cases in the multivariate case is more complicated.

A theoretical justification for the unsatisfactory behaviour of the maximum likelihood 
estimates is that the symmetrized Kullback-Leibler divergence between two $SN_p$ densities
with similar values of $\balpha$ tends to be very small; this fact typically produces a
profile likelihood for $\balpha$ which is rather flat over a large portion of the
parameter space.
Another way of interpreting the difficulties of a likelihood approach, at least in 
a simple setting, is the following. For a fixed positive value $z$, consider, as a
function of $\alpha$, the likelihood ratio between a standard normal density and an
$SN(0,1,\alpha)$ one, with positive $\alpha$, that is 
$$
LR(\alpha) = \frac{\varphi(z)}{2 \varphi(z) \bPhi(\alpha z)}= \frac{1}{2\bPhi(\alpha z)}.
$$
Since $LR(\alpha)$ is decreasing, for any fixed positive $z$, in $(0,\infty)$, its 
possible values range from $0.5$ (when $\alpha\rightarrow \infty$, that is for a
half-normal density) to 1 (for $\alpha=0$); in other words the ability of the likelihood
to discriminate between a normal and a skew-normal model seems quite limited.
One possibility is then to switch to the production of 
valid interval estimates.
However, solid classical and likelihood theories of confidence intervals for the $SN_p$
model are still lacking.
Another technical inferential problem with the SN model is that the 
likelihood function may be multimodal when both the location and the shape parameters
are unknown: we will discuss this issue below in this section.  

For all these reasons, we propose a full Bayesian analysis of the
multivariate SN model.
A Bayesian analysis based on objective priors has already been proposed by \citet{lisl:05} for the scalar case.
See also \citet{pew:08} for an objective Bayesian analysis in the half-normal and half-$t$ cases, and \cite{Bra:10}
for the skew-$t$ case.
\cite{sylv:10} have recently proposed a fully Bayesian analysis of a mixture of
skew-normal and skew-$t$ densities.
Other recent and important advances in the application of multivariate skew-normal models can be found in \cite{fs:10},
\cite{ps:10}, \cite{FerMou12} and \cite{CabLac12}.\\
The computational approach described in \cite{sylv:10} differs from ours in two respects: 
i) they adopt conjugate priors in order to facilitate a Gibbs sampling strategy for simulating from the posterior; ii) 
as a consequence of i), we adopt a different sampling strategy, based on importance sampling rather than MCMC; 
we will describe the PMC algorithm in detail in Section \ref{Population Monte Carlo algorithm}.
For the moment we explain why we are not completely confident with the use of Gibbs-type algorithms
for skew-normal or skew-$t$
models. It is a well-known fact that likelihood functions arising from a skew-normal model
may be multimodal. In these situations,
the Gibbs sampler chains are often captured by one of the modes.
As a consequence, the chains do not mix well and the posterior distribution is not well explored.

\begin{figure}[ht]
 \centerline{\includegraphics[width=\textwidth,width=0.7\textwidth]{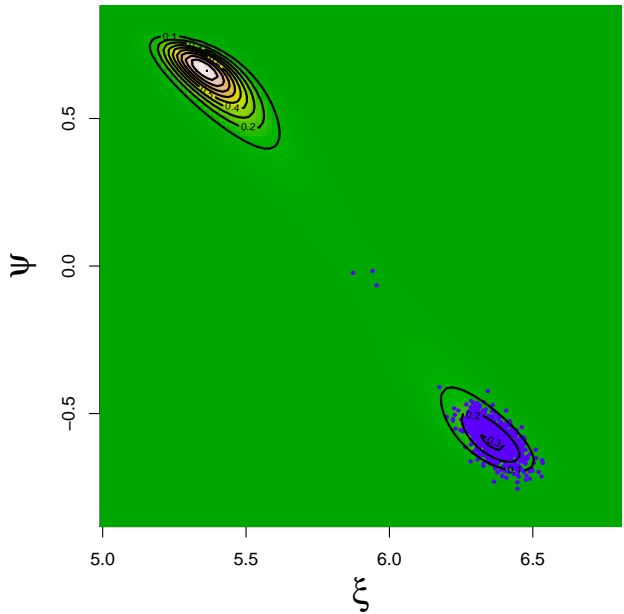}} 
 \caption{Scalar skew-normal example; posterior distribution for $(\xi, \psi)$ - where $\psi=\omega \alpha/\sqrt{1+\alpha^2}$ 
 and 1000 values drawn from a Gibbs algorithm.}
 \label{twomodes}
\end{figure}

As a practical illustration of the problem, Figure \ref{twomodes} presents
the 1000 draws obtained from a Gibbs sampler - 
similar to that proposed in \cite{sylv:10} -
in the very simple setting of a scalar skew-normal model with unknown location $\xi$ and
shape $\alpha$, and a known scale parameter 
$\omega=1$. 
Almost all posterior draws belong to the same mode and the posterior distribution is not
well explored. 
In the multidimensional case, things tend to be more complicated;
as we will argue in Section \ref{22}, constraints in the
parameter space should be introduced in order to obtain a positive definite correlation matrix, 
and accounting for them in the Gibbs sampling algorithm may not be easy.   

\section{Augmented likelihood function and priors}
\subsection{Introducing the latent structure}
\label{22}
In this section we describe how to exploit the intrinsically latent structure of the 
SN density function in order 
to produce an augmented likelihood function. 
The main proposition follows.
\begin{propo}
Let $\bOmega$ be a correlation matrix, $\bdelta$ a $p$-dimensional vector and 
${ \balpha}=(1-\bdelta^T\bOmega^{-1}\bdelta)^{-\mez}\bOmega^{-1}\bdelta$. Define
$$
\binom{Z}{\bX} \sim  N_{p+1}
\left [ \binom{0}{\bzero} , 
\left ( 
\begin{array}{c c}
1    &\bdelta^T \\
\bdelta  & \bOmega 
\end{array}
 \right)
 \right ]
\mbox{ and }
\bU= \begin{cases}
\bX & Z \geq 0 \\
-\bX & Z < 0
\end{cases}.
$$
Then, (a) the random vector ${ \bY= \bomega \bU + \bxi} \sim SN_p(\bSigma, \bxi, \balpha)$, with $\bSigma=\bomega \bOmega \bomega,$ 
and (b) the joint density of $(\bY, Z)$ is given by
$$
f_{p+1}\left ( \by,z\right ) = f_p(\by\mid z) f(z)
= { N_p\left (\bxi + \bomega \bdelta \mid z \mid, \bomega (\bOmega-\bdelta\bdelta^\prime)\bomega \right )}\, \times N_1(0,1).
$$
\end{propo}

\noindent
{\bf Proof:} (a): From one of the possible definitions of a multivariate $SN$ r.v., it is known that $\bU\sim SN_p(\bOmega, 
\bzero, \balpha)$; $\bY$ is a simple linear transformation of $\bU$ and its distribution is readily obtained.\\
(b): Start from $f(y,z) = f(z) f(y\mid z)$. Then $f(z)$ is, by assumption, a standard Gaussian density, while
$$
\left (\bY \mid Z=z \right )= \left (\bomega \bU + \bxi \mid Z=z \right )= 
\begin{cases}
\bomega \bX + \bxi & z \geq 0 \\
-\bomega \bX + \bxi & z < 0 \\
\end{cases}.
$$
Then, by using simple results on conditional Gaussian densities, one gets
$$
\left (\bY \mid Z=z \right )\sim 
\begin{cases}
N_p\left (\bxi + \bomega \bdelta z, \bomega (\bOmega - \bdelta\bdelta^\prime) \bomega\right ) & z \geq 0 \\
N_p\left (\bxi - \bomega \bdelta z, \bomega (\bOmega - \bdelta\bdelta^\prime) \bomega\right ) & z < 0 \\
\end{cases}. 
$$
Using the above proposition  one can write an augmented likelihood function, ``as if'' we
had observed, for each sample unit, 
the latent value $z_i$, $i=1, \dots, n$. Write 
$\bpsi= \bomega \bdelta$ and 
$\bomega (\bOmega-\bdelta\bdelta^\prime)\bomega ={ \bSigma - \bpsi\bpsi^\prime}=\bG;$ 
define the parameter vector as $\bte^\ast=(\bdelta, \bSigma, \bxi)$ or
$\bte=(\bpsi,\bG, \bxi)$. The augmented likelihood function is then  
\begin{eqnarray*}
{L(\bte; \by, \bz)} &\propto&   
\prod_{i=1}^n \left \{ \varphi_p(\by_i-\bxi-\bpsi\,\mid z_i\mid; \bSigma-\bpsi\bpsi^\prime) \times  
\varphi_1(z_i;1)\right \} \\
&=& {\frac 1{\mid \bG \mid^{\frac n2}} \exp\left(-\frac 12 \sum_{i=1}^n z_i^2\right ) } 
 \\ &\times&\exp\left (-\frac 12 \sum_{i=1}^n (\by_i-\bxi-\bpsi \mid z_i\mid)^\prime 
 \bG^{-1}  (\by_i-\bxi-\bpsi \mid z_i\mid)\right ).
\end{eqnarray*}
Notice that the matrix $\bG$ must be {positive definite}; this implies a logical
constraint among the values of $\bdelta$ and $\bOmega$ in the original parameterization 
which should be taken into account 
when exploring the parameter space via simulation methods. 
As we have already noticed, this issue seems to
have been neglected in literature. See
Azzalini's website \texttt{http://azzalini.stat.unipd.it/SN/}, under the section ``A
less frequent question'' for a graphical treatment of this problem.
In particular, MCMC methods should be used with care in order to avoid the chain 
in the $(\bdelta,\bOmega)$ parameterization visiting inadmissible parts of the
parameter space.

\subsection{Prior distributions}
\label{Prior distributions}
Our primary goal here is to propose a general method of inference for the parameters of the multivariate SN distribution.
For these reasons we have tried to be as ``objective'' as possible in choosing the prior for the parameter vector.
However, it is not easy to derive a formal Jeffreys or reference prior for the parameters of a multivariate skew-normal distribution.
In this paper we have assumed a priori, as usual, $\bxi \independent (\bdelta,\bSigma)$. 
Also we have assumed a flat prior for the ``location'' parameter $\bxi$ and a conjugate normal 
Inverse Wishart prior for the ``scale'' parameter $\bSigma$, that is
$$\pi(\bxi) \propto 1 \qquad \mbox{  and } \qquad  
{\bSigma \sim IW_p(m, \bLambda)}. 
$$
Obviously, one can always consider the limiting case ($m\rightarrow 0, \bLambda \rightarrow \bzero$) 
to get the classical Jeffreys prior
\beq
\label{ref_sigma}
\pi(\bxi, \bSigma) \propto \frac 1{\mid \bSigma \mid^{\frac{p+1}{2}}}.
\enq
The choice of a good objective prior for $\bdelta$ (or $\balpha$) is more delicate. 
 \cite{lisl:05} have shown that, in the univariate SN model, the Jeffreys' prior for the shape parameter $\alpha$ is proper; 
its use, in a sense, automatically and pragmatically solves the problem of a potentially non vanishing likelihood function, which can 
happen with the skew-normal model \citep{ac:99}.
 \cite{bra:05} have shown that the Jeffreys' prior can be adequately approximated by a Student $t$ density with a half 
 degree of freedom, centered at zero and with scale parameter $\pi^2/4$.
\cite{Bratc:10}, in an as yet unpublished technical report, have partially generalised 
the above results to the bivariate case, but no general results are available for the $SN_p$ model with $p>2$.
They have proved that, unlike the scalar case, the Jeffreys' prior is improper in the bivariate case. 
On the other hand the one-at-a-time reference prior \citep{bebe:92} is proper although its expression is quite complicated.
In particular, the Jeffreys' prior in the $\balpha$  parameterization (using the same approximation provided by \cite{bra:05}
for the scalar case) is
$$
\pi^J(\alpha_1, \alpha_2) \propto \frac 1{1+ \pi^2/2 (\alpha_1^2+ \alpha_2^2)}.
$$
The {\em proper} reference prior when $\alpha_1$ is the parameter of interest and $\alpha_2$ is 
considered a nuisance parameter is given by $\pi^R(\alpha_2\mid \alpha_1)\pi^R(\alpha_1)$ where 
\beq
\label{duedatouno}
\pi^R(\alpha_2\mid \alpha_1)   \propto
\frac {\left (1+ 2\eta^2 \alpha_1^2\right )^{1/4}}
{\left (1+ 2\eta^2 (\alpha_1^2+\alpha_2^2)\right )^{3/4} }
\frac 1{\sqrt{\left (1+ 2\eta^2\alpha_1^2\right )}}
\enq 
and 
\beq
\label{uno.marg}
\pi^R(\alpha_1) \propto
\exp
\left ( -\frac 14 \int \log \left ( 1+ 2\eta^2 (\alpha_1^2+ t^2)\right )\pi_R(t\mid \alpha_1)d t 
\right ),
\enq
 where $\eta=\pi/2$. 
 Of course, when $\alpha_2$ is the parameter of interest the prior is the same 
with $\alpha_1$ and $\alpha_2$ switching their roles.
The above considerations show that an objective analysis can be made only for one 
component of the shape vector: to get a proper posterior with sampling probability $1$, in
the multivariate case, one should introduce genuine prior information for some of the
components of $\balpha$. For practical purposes,
a prior can be chosen in the following way: in the scalar case the approximate 
Jeffreys' prior for $\beta=(1+\delta)/2$, 
with  $\delta=\alpha/\sqrt{1+\alpha^2}$, is a Beta$(0.25,0.25)$ prior;
in analogy with that, one can use, in the multivariate case, the prior
\beq
\label{appr}
\pi^{IND}(\bdelta) \propto \prod_{j=1}^p \left ( 1-\delta_j^2\right )^{-\frac 34},
\enq
that is, we assume that the components of the skewness vector are, a priori, independent 
and identically distributed. Although independence can be considered a strong assumption,
it is hard to conceive any non subjective form of dependence. 
Alternatively, the use of a uniform prior in the $\bdelta$ parameterization, especially 
for $p>2$, could be suggested. 
Using the Jacobian 
$$
\left | \frac{\partial \bdelta}{\partial \balpha} \right | = \frac{\mid \bOmega \mid}{\left ( 1 + \balpha^\prime \bOmega \balpha\right )^2},
$$ 
the uniform prior in the $\bdelta$ parameterization is transformed into 
\beq
\label{uniform}
\pi^U(\balpha) \propto
\frac 1{(1 + \balpha^\prime \bOmega \balpha)^2},
\enq
which explicitly introduces a dependence on the correlation matrix $\bOmega$.
Notice also that any prior distribution on $\bdelta$ should be considered only 
for those values which satisfy the constraints illustrated at 
the end of section \ref{22}.
In this perspective, for computational reasons, we will consider the parameter constraint
as 
generated from the prior rather than from the likelihood. 
In the rest of the paper, we will then consider, as the prior for $\delta$, 
\beq
\label{appr2}
\tilde{\pi}^{IND}(\bdelta\mid \bOmega) = \frac 1{A(\bOmega)} \prod_{j=1}^p \left ( 1-\delta_j^2\right )^{-\frac 34},
\enq
where 
$A(\bOmega)$ is the integral of (\ref{appr}) over the parameter values such that
$\det(\bG)>0$.
In fact, it is important to notice that, given the hierarchical structure of the prior for
$\bdelta$, 
one needs to recover the normalizing constant $A(\bOmega)$. This is analitically feasible only
for particular choices of 
$\pi(\bdelta\mid \bOmega)$. In all other cases, one needs to evaluate the integral of
$\pi(\bdelta\mid \bOmega)$ on the ellipsoid determined by the constraint.

\begin{figure}
 \begin{center}
  \includegraphics[width=\textwidth]{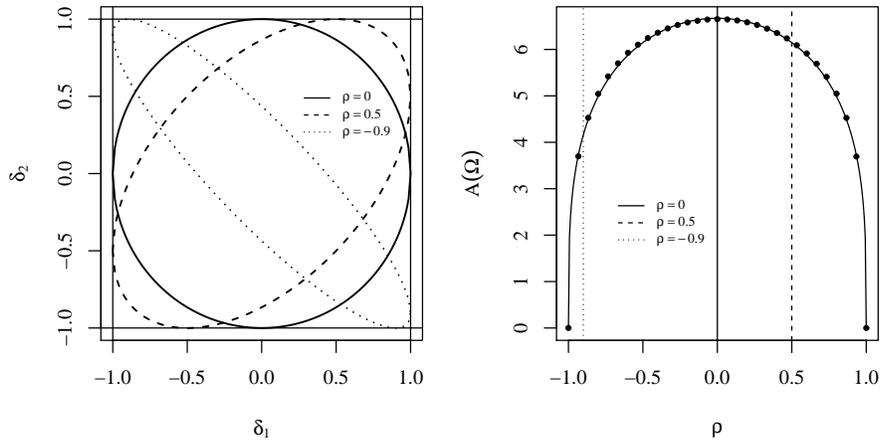}
 \end{center}
 \caption{Left panel: different shapes of the parameter space of $\bdelta$ for different
values of $\rho$. 
Right panel: dots represent numerical evaluations of $A(\bOmega)$ for different values of
$\rho$, the solid curve represents the approximation given in eq.\,(\ref{eq:approx}).
Vertical lines denote the same values of $\rho$ considered in the left panel.}
 \label{fig:ConstEll}
\end{figure}

We now give some details about the evaluation of $A(\bOmega)$ when $p=2$.
Generalizations to higher dimensions are similar, although more complicated. 
In a bivariate setup, we define $\rho$ to be the off-diagonal element of $\bOmega$; 
the constraint produces an ellipse which is a proper subset of the square $[-1,1]^2$ (see figure
\ref{fig:ConstEll}, left panel): 
the shape of the ellipse is a function of $\rho$.
$A(\bOmega)$ can be evaluated on a grid of values of $\rho$, for example using  
a rejection sampler where the simulated values of $\bdelta$ are 
generated by two independent $\textrm{Beta}({0.1,0.1})$ deviates. The choice of the
proposal is due to the fact that the Jeffreys' prior puts most of the probability mass on
the boundary of the square. Results on a grid of values of $\rho$ are represented in the right panel of 
fig. \ref{fig:ConstEll} by dots.
This approach can be computationally demanding. For practical purposes, a very satisfactory approximation 
can be obtained using the formula
\begin{equation}\label{eq:approx}
 A(\bOmega) \approx a\,(1-\rho^2)^b.
\end{equation}
Estimation of $a$ and $b$ is then straightforward. We have obtained $a$ = 6.68 and $b$ = 0.28. 

Although the above defined $\btheta^\ast$ parameterization is more suitable for elicitation, 
the alternative $\btheta$ parameterization
should be preferred in terms of implementation and computation. From now on, 
we will use $\btheta=(\bpsi, \bG, \bxi)$.
This can be simply done by introducing a Jacobian term in the prior, namely
\begin{equation}
 \label{jac}
J(\btheta^\ast \to \btheta) = \prod_{j=1}^p \left ( G_{jj} + \psi_j^2 \right )^{-\frac 12}.
\end{equation}

\section{Population Monte Carlo algorithm}
\label{Population Monte Carlo algorithm}

In this section we illustrate a PMC algorithm for obtaining a sample from the joint
posterior 
distribution  of $\btheta$.
PMC methods (see e.g. \citealp{MR2109057}) essentially consist of an iterated 
version of the importance sampling algorithm: 
at each iteration, a population of particles is generated, independently of each other, 
possibly using a set of different importance functions. 
Performances obtained in the past iterations by the different kernels are typically 
evaluated in a relative way in order to adaptively 
modify the proposal distributions over the iterations.\\
Alternatively, \citet{MR2236856} suggest the use of the full conditional distributions as
the importance functions when the model at hand has a latent structure representation, as
in the present case.
This way, one can exploit the easiness of proposing from a natural importance function,
i.e. the full conditional, and, at the same time, avoid the convergence issues of a
generic MCMC method. Also,
the coexistence of different particles, and the competition between them, 
allows us to tackle better the issue of multimodality of the posterior density. 
It is well known that in similar situations
the Gibbs sampler tends to be attracted by 
one of the modes and hardly escapes from a neighborhood of it \citep{celeux:00}.

From a model selection perspective, the estimation of the normalising 
constant of $\pi(\btheta\vert \by)$
can be performed as a simple by-product of any PMC (and MC) sampler. 
In fact, from the importance sampling identity, one obtains
\begin{displaymath} 
p(\by)=\int_\bTheta \int_{\mathcal Z}\pi(\btheta,\bz,\by)d\btheta d\bz =
\int_\bTheta\int_{\mathcal Z} \frac{\pi(\btheta,\bz,\by)}{q(\btheta,\bz)}q(\btheta, \bz)
d\btheta d\bz,
\end{displaymath}
where $q$ is the proposal distribution. Adopting the usual Monte\,Carlo approximation,
$p(\by)$ can be estimated by
\beq
\hat{p}(\by) \approx \frac{\sum_{t=1}^T H_t \sum_{i=1}^N \tilde \zeta_i^{(t)}}{N\sum_{t=1}^T H_t},
\label{bf}
\enq
where the $\tilde{\zeta}_i$'s denote the unnormalised importance weights, and
\begin{equation}
\label{ht}
 H_t = -\sum_{i=1}^N \zeta_i^{(t)}\log(\zeta_i^{(t)}).
\end{equation}
is an entropy measure of performance of the $t$-th iteration of the algorithm. 
$H_t$ takes high values when the normalised weights of the particles in the $t$-th iteration 
are concentrated around $1/N$.
The quantity (\ref{ht}) is a monotonic transformation of the perplexity measure 
%the \textrm{Essential Sample Size} and the \textrm{Perplexity}
\citep{RoCa2010}, defined as $\exp(H_t)/N$.
We will use the perplexity index as a measure of non degeneracy of the PMC algorithm, which 
is often considered as one potential drawback of Monte\,Carlo methods. 
Last but not least, 
the use of PMC algorithms allows the simultaenous draw of all the particles: 
this fact dramatically improves the efficiency 
of the algorithm compared with generic MCMC approaches.
The estimator (\ref{bf}) is quite simple and stable since it does not rely upon
convergence issues.  
A possible improvement on (\ref{bf}) can be obtained via the Adaptive Multiple Importance sampling technique, 
\cite{AMIS}.
 The difference with PMC is that, in this case, the importance weights of all simulated values, 
 past as well as present, are recomputed at each iteration. We are currently working on these aspects.

Without loss of generality, we illustrate the steps of the algorithm for a bidimensional setup. The generalization 
to higher dimensional problems is straightforward, even though numerical problems can arise with some choices of the 
prior for $\bdelta$, and care must be taken in handling the approximation of $A(\bOmega)$.
In the simulation study described in Section \ref{Examples} we reported perplexities of
our samples, although we have never experienced significant problems in terms of
degeneracy: however we recognize that this can be a critical issue when the proposal
densities are not well calibrated.

\begin{figure}[htb]
 \centering
 \includegraphics[width=0.8\textwidth]{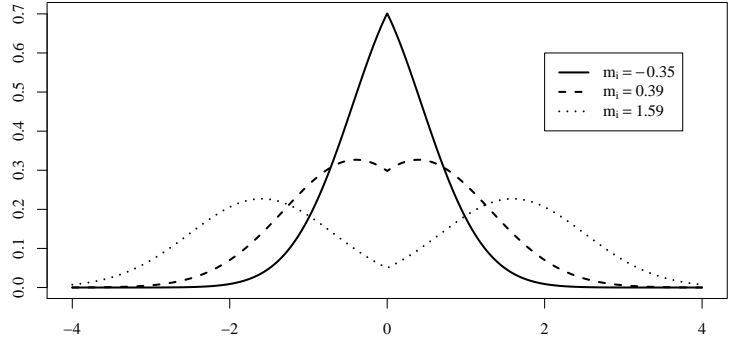}
 \caption{Some examples of the full conditional distribution of a generic $z_i$ for different values of $m_i$.}
 \label{fig:Densita2}
\end{figure}

After approximating $A(\bOmega)$, 
the PMC algorithm can be initialised by sampling random starting particles. These
particles will be updated in the following iterations using, as proposal distributions, 
the full conditionals (when available in closed form) or some other distributions which approximate them.\\
In particular, the full conditional distributions of the latent variables $Z_i$'s (see figure \ref{fig:Densita2}) 
are symmetric about the origin:
\begin{displaymath}
 f(z_i|\cdots)=\left\{
 \begin{array}{lll}
  \varphi(z_i-m_i;v)& \rule{0.2cm}{0cm}& z_i\geq 0\\
  \varphi(z_i+m_i;v)& \rule{0.2cm}{0cm}& z_i< 0\\
 \end{array}\right. ,
\end{displaymath}
where
\begin{displaymath}
 v = (1+\bpsi^\prime \bG^{-1}\bpsi)^{-1},
\end{displaymath}
and $m_i$ is the $i$-th component of the vector 
\begin{displaymath}
  \bm = v \left[(\by -\mathds 1_n \otimes \bxi)^\prime \bG^{-1}\bpsi\right].
\end{displaymath}
It is not necessary to sample the signs of $Z_i$'s, as $\bz$ is involved in the
posterior distribution 
only with its absolute value, and the full conditional distribution of $|z_i|$ is just a
truncated normal. Nevertheless, we prefer to directly sample the value of $\bz$, as
potential asymmetries in the posterior draws could highlight potential problems in the sampler.
Hence, we only have to sample the sign (each sign having probability 1/2) independently of
$|z_i|$. Generation of the $z_i$'s  can be done using several methods; troubles can arise
when $m_i/v^{1/2}$ takes large negative values: in this case, sampling from the very
extreme tail of a normal distribution using an accept-reject algorithm can be 
intensive while the inversion method may give numerically unreliable results; in these
cases we have employed the approach described in \cite{MR1963334}, essentially a  perfect
sampling algorithm.

Simple algebra leads to the full conditional for $\xi$:
\begin{displaymath}
  \left [ \bxi|\cdots \right ] \sim N_p (\bar \by-\bpsi\overline{|\bz|}, \bG/n),
\end{displaymath}
where $\bar \by$ is the sample mean vector, and $\overline{|\bz|}$ is the mean of the 
absolute values of the $z_i$'s.
Finally, $\bG$ and $\bpsi$ have non-standard full-conditional distributions:
\begin{displaymath}
  \pi(\bG|\cdots) \propto \pi(\bG) IW_p(n+m,\bLambda_\star),
\end{displaymath}
where $\pi(\bG)$ is the prior for $\bG$ arising from (\ref{ref_sigma}) and a Jacobian argument, and 
\begin{displaymath}
  \bLambda_\star=\bLambda + \sum_{i=1}^n (\by_i-\bpsi |z_i|-\bxi)(\by_i-\bpsi |z_i|-\bxi)'.
\end{displaymath}
We use a $IW_p(n+m,\bLambda_\star)$ distribution (which is the one obtained by the full 
conditional distribution ignoring the contribution of the prior) to propose values for
$\bG$, as this distribution will resemble the full conditional, in particular for large
sample sizes.
Finally, the full conditional distribution of $\bpsi$ is proportional to
\begin{displaymath}
 \pi(\bpsi|\cdots) \propto \pi(\bpsi| \bOmega) \varphi_p \left( \bpsi -  
\frac{\sum_{i=1}^n |z_i|(\by_i-\bxi)}{\sum_{i=1}^n z_i^2}; \frac{\bG}{\sum_{i=1}^n
z_i^2}\right ).
\end{displaymath}
In this case, it is possible to consider several proposal distributions; 
in order to minimize the computational burden, we propose to sample values from the
$p$-variate normal ``part'' of the full conditional distribution. 

As usual, we compute the importance weights $\tilde{\boldsymbol{\zeta}}$ as the ratio
$\tilde\pi(\btheta,\bz\mid\by)/q(\btheta,\bz)$, where $\tilde\pi$ is the posterior density in which
the prior for $(\bpsi\mid\bOmega)$  has been (approximately) normalised, and $q$ is the joint
proposal density. Particles will be multinomially resampled with unnormalised weights
given by $\tilde{\boldsymbol{\zeta}}$ and the resampled particles will represent the starting point for
the particles of the next iteration.

\section{Bayes factor}
\label{4}

One of the main reasons for the popularity of the multivariate skew-normal model
is that it represents a proper generalization of the multivariate normal model. Then it 
is often important to test the normality of the dataset versus skew-normal
alternatives. Here we will use the Bayes factor to compare the multivariate Gaussian
model - say $M_0$ - versus the multivariate skew-normal one - say $M_1$. 
To this end, we need to evaluate the
predictive distribution of the data under the two competing models. 
Suppose that, under model $M_0$, $\bpsi$ is set equal 
to $\bzero$. 
Then the model is described by the following assumptions:
\bei 
\item $Y_1, \dots, Y_n \iid N_p(\bxi, \bSigma)$;
\item $\pi^J(\bxi, \bSigma) = \mid \bSigma\mid^{-(p+1)/2}$;
\eni
it is a standard calculation to show that the marginal distribution of the data under
the normal model is 
\beq
\label{mzero}
p_0(\by) = \int_{\bSigma}\int_{\bxi} \varphi_p(\by-\bxi; \bSigma) \pi^J(\bxi, \bSigma) d\bSigma 
d\bxi = \frac{2^{n/2} \Psi_p((n-1)/2)}{\pi^{p(n-p-1)/4} \mid S \mid^{(n-1)/2} n^{n/2}},
\enq
where $S$ is the sample covariance matrix and $$ \Psi_p(u) = \prod_{j=1}^p \Gamma\left (u - \frac 12(j-1)\right )
$$
is the multivariate Gamma function.
Notice that, since the Jeffreys' prior 
$\pi^J(\bxi, \bSigma)$ is improper, quantity (\ref{mzero}) is meaningless {\it per se}.
However, if we use the same - improper - prior for the common parameters of the two models
(in this case $\bxi$ and $\bSigma$), then the Bayes factor is a well-defined tool for
model comparison.

\noindent
To compute the Bayes factor $B_{10}$ 
for comparing the skew-normal model and the nested normal model we need an estimate $\hat p_1(\by)$ 
of $p_1(\by)$, the marginal distribution 
of data under the skew-normal hypothesis.   
We then perform $T$ iterations of PMC algorithm and we sample $N$ particles in each iteration.
Using (\ref{bf}), the final estimate of the Bayes factor is then 
\begin{equation}
 B_{10}\approx \frac{\sum_{t=1}^T H_t \sum_{i=1}^N \tilde \zeta_j^{(t)}}{ p_0(\by) N\sum_{t=1}^T H_t}.
\label{bbff}
\end{equation}

\section{Some discussion and extensions}
\label{5}
The computational approach we have discussed in the previous sections can be easily adapted to more general 
situations. In the presence of $k$ covariates, the location parameter vector $\bxi$ should be replaced by a $k \times p$ 
matrix of regression coefficients $\bB$, so that our model gets transformed into 
\begin{displaymath}
 \by_i \iid SN(\bG, \bX_i' \bB,\bpsi),\qquad i=1,2,\ldots,n.
\end{displaymath}
The augmented likelihood for this new model is then 
\begin{eqnarray*}
 L(\bB, \bG,\bpsi|\bY, \bz) & \propto &|\bG|^{-n/2} \exp\left\{-\frac 1 2 \bz' \bz\right\}\\
 & &\exp\left\{-\frac 1 2 \sum_{i=1}^n (\by_i-\bX_i' \bB-\bpsi |z_i|)' \bG^{-1}
(\by_i-\bX_i' \bB-\bpsi |z_i|)\right\}.
\end{eqnarray*}
The previous PMC sampler is still valid for this model; the only necessary modification is
the introduction 
of a proposal step for 
$\bB$ in lieu of $\bxi$. Adopting a flat prior for the elements of the matrix $\bB$, we
again use the full conditional 
distribution of 
$\bB$ as our proposal. It is easy to show that 
\begin{displaymath}
 [\bB|\cdots] \sim MN(\bS^{-1}\bC_{\bpsi}, \bS^{-1}, \bG),
\end{displaymath}
where
\begin{displaymath}
 \bS = \sum_{i=1}^n \bX_i \bX_i', \qquad \bC_{\bpsi} =\sum_{i=1}^n \bX_i (\by_i -
|z_i|\bpsi)',
\end{displaymath}
and the symbol $MN(\bM, \bR, \bDelta)$ refers to a matrix normal random variable $\bV$
\citep{Da81}, 
with location $\bM$ and scale parameters $\bR$ and $\bDelta$, with density:
\begin{displaymath}
 f(\bV|\bM,\bR,\bDelta) = \frac{\exp\left\{\frac 1 2 \operatorname{tr}[\bR^{-1}
(\bV-\bM)'\bDelta^{-1}(\bV-\bM)]\right\}}{(2\pi)^{np/2} |\bR|^{n/2} 
|\bDelta|^{p/2}}.
\end{displaymath}
Simulating draws from this distribution is simple, as it is linked with the multivariate
normal 
distribution by a simple relation:
\begin{displaymath}
 \bV\sim MN_{n\times p}(\bM, \bR, \bDelta)\mbox{ if and only if }\operatorname{vec}(\bV)
\sim N_{np}(\operatorname{vec}(\bM), \bR\otimes\bDelta),
\end{displaymath}
where $\otimes$ denotes the Kronecker product.

The Bayesian approach through data augmentation is also particularly useful in problems
with missing data. Our algorithm can be easily adapted to account for missingness and a
comparison of our approach with the one based on the EM algorithm proposed in \cite{Lin1}
and \cite{Lin2} is currently under investigation.

\section{Simulations and examples}
\label{Examples}
In this section we consider the frequentist properties of our Bayesian procedure; 
in particular we simulate samples of size $n_1=50$ and $n_2=200$ for different
combinations of parameter values. In
all simulations we have used $20,000$ particles for 20 iterations, setting
$\bxi=(3,3)^\prime$ and $\omega_1=\omega_2=1$.\\
Table \ref{tab:results} shows a summary of the results: for each parameter combination 
we provide
\begin{itemize}
\item the frequency of times that $B_{10}$ provides evidence in favour of the normal or 
the skew-normal model (second and fourth column) or that it does not provide strong evidence for
any of the two models (third column);
 \item the median of the simulated sampling distribution of the posterior median 
(fifth column);
 \item the frequentist coverage of the one-sided 0.95\% and 0.9\% credible sets 
(columns $FC_{0.95}$ and $FC_{0.9}$);
 \item the median of the sampling distributions of the posterior 
mean $\expe{\psi_1|\by}$ (eighth column, denoted by MeMean), Conditional (on the true $z_i$'s) MLE (ninth column,
denoted by MeCMLE) and MLE (tenth column, denoted by MeMLE);
\end{itemize}
Results for the skewness parameter are shown for the first component 
of the vector $\bpsi_1$; similar results are obtained for $\bpsi_2$.
As might be expected, both the likelihood and Bayesian approaches successfully
estimate the off-diagonal element of $\bG$, while estimation of $\bpsi$ - and,
consequently, of $\bxi$ - seems more difficult. Table  \ref{tab:results} highlights the difficulties in
catching skewness in small datasets, which implies a high rate of wrong answers
given by Bayes factors for non-normal samples.
With respect to the ML estimator, it should be noticed that, even though the medians of the 
sampling distributions of the MLE are quite
precise, this estimator always shows a non-negligible probability of producing
infinite estimates.\\
\begin{sidewaystable}
 \begin{small}
    \begin{displaymath}
   \begin{array}{cccccccccccc}
    n=30\\
    \multicolumn{3}{c}{\text{True values}}&B_{10}<0.5&0.5\leq B_{10}<2&B_{10}\geq 2&Med(Med(\psi_{1}))&FC_{0.95}&FC_{0.9}&MeMean&MeCMLE&MeMLE\\
    \hline
    \bpsi=0_2,&\rho=0&(\balpha=0_2)& 0.997&0.003&0.000&0.007&0.983&0.971&0.008&0.010&-0.015\\
    \bpsi=0.5_2,&\rho=0&(\balpha\approx 0.71_2)& 0.988&0.011&0.001&0.031&0.926&0.874&0.026&0.516&0.085\\
    \rowcolor[rgb]{0.8,0.8,0.8}\bpsi=0.7_2,&\rho=0&(\balpha\approx 4.95_2)& 0.877&0.098&0.025&0.362&0.806&0.728&0.308&0.696&0.582\\
    \bpsi=0_2,&\rho=0.5&(\balpha=0_2)& 0.996&0.003&0.001&0.011&0.979&0.959&0.007&0.022&0.123\\
    \bpsi=0.5_2,&\rho=0.5&(\balpha\approx 0.41_2)& 0.989&0.009&0.002&0.031&0.900&0.852&0.026&0.519&0.146\\
    \bpsi=0.7_2,&\rho=0.5&(\balpha\approx 0.79_2)& 0.940&0.042&0.018&0.163&0.782&0.660&0.133&0.702&0.385\\
    \rowcolor[rgb]{0.8,0.8,0.8}\bpsi\approx 0.495_2,&\rho=-0.5&(\balpha=7_2)& 0.958&0.034&0.008&0.161&0.874&0.838&0.116&0.499&0.483\\% delta = 0.4949747
    &&&&&\\
    &&&&&\\
    n=200\\
    \multicolumn{3}{c}{\text{True values}}&B_{10}<0.5&0.5\leq B_{10}<2&B_{10}\geq 2&Med(Med(\psi_{1}))&FC_{0.95}&FC_{0.9}&MeMean&MeCMLE&MeMLE\\
    \hline
    \bpsi=0_2,&\rho=0&(\balpha=0_2)& 0.999&0.001&0&-0.002&0.873&0.816&-0.003&-0.007&0.045\\
    \bpsi=0.5_2,&\rho=0&(\balpha\approx 0.71_2)& 0.995&0.004&0.001&0.211&0.709&0.634&0.187&0.500&0.267\\
    \rowcolor[rgb]{0.8,0.8,0.8}\bpsi=0.7_2,&\rho=0&(\balpha\approx 4.95_2)& 0.006&0.005&0.989&0.573&0.737&0.635&0.569&0.699&0.687\\
    \bpsi=0_2,&\rho=0.5&(\balpha=0_2)& 0.999&0.001&0.000&0.027&0.820&0.724&0.031&-0.005&0.071\\
    \bpsi=0.5_2,&\rho=0.5&(\balpha\approx 0.41_2)& 0.997&0.002&0.001&0.152&0.695&0.601&0.139&0.499&0.183\\
    \bpsi=0.7_2,&\rho=0.5&(\balpha\approx 0.79_2)& 0.899&0.065&0.036&0.479&0.701&0.624&0.460&0.699&0.664\\
    \rowcolor[rgb]{0.8,0.8,0.8}\bpsi\approx 0.495_2,&\rho=-0.5&(\balpha=7_2)& 0.008&0.019&0.973&0.385&0.885&0.804&0.382&0.496&0.495\\% delta = 0.4949747
   \end{array}
  \end{displaymath}
 \end{small}
 \caption{Summary of the results obtained in the simulation study; the rows relative to 
combinations of parameters that generate high skewness are highlighted in gray.}
\label{tab:results}
\end{sidewaystable}
\begin{figure}
 \includegraphics[width=\textwidth]{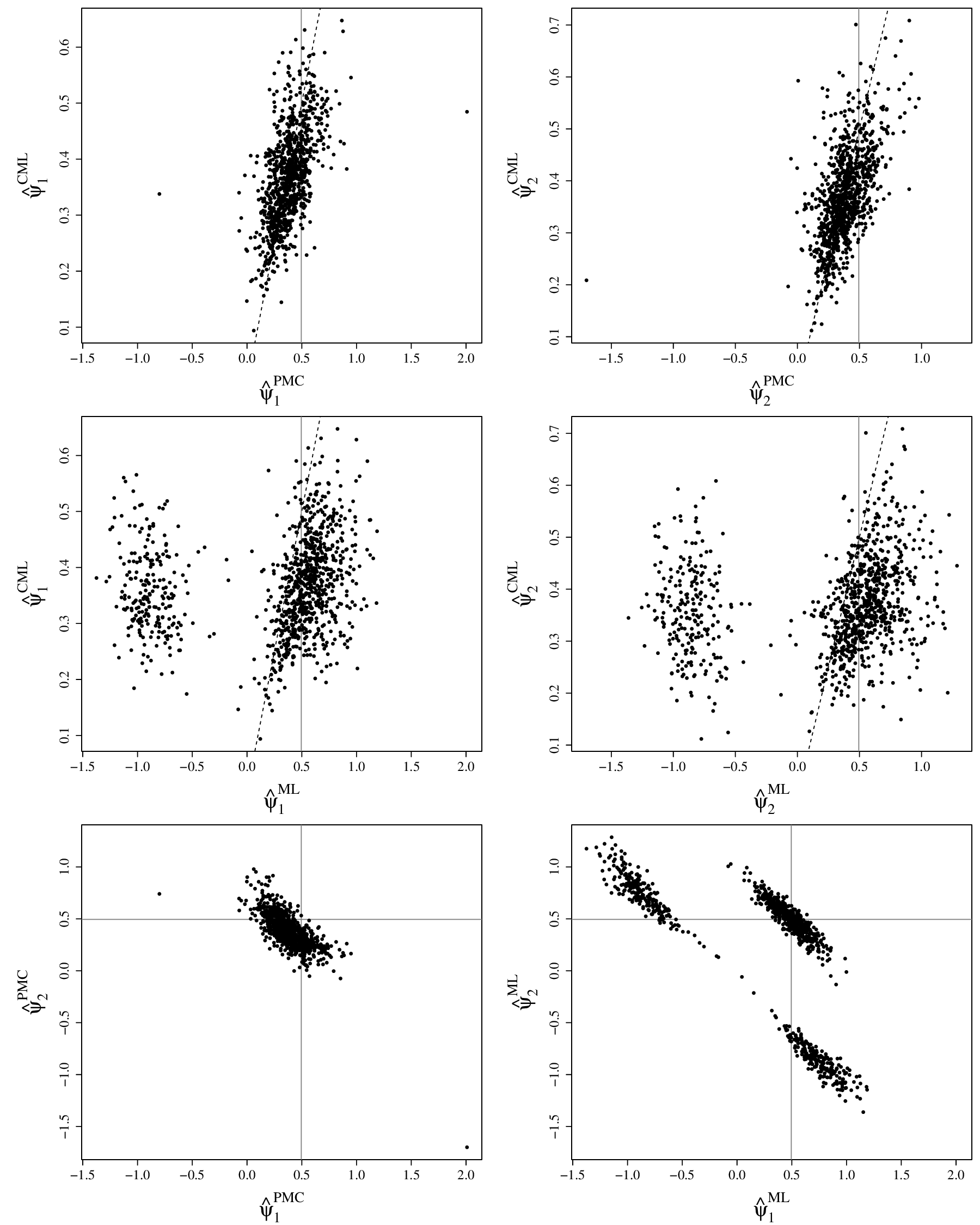}
 \caption{ML, CML, Bayesian estimates and true values of the parameters.}
 \label{fig:Confronti}
\end{figure}
Figure \ref{fig:Confronti} compares the estimates obtained using our approach
and the maximum likelihood method in the most extreme combination of parameters, with
$\rho=-0.5$, $\bomega=(1,1)^\prime$ and $\bpsi=(0.495,0.495)^\prime$. The value of
$\bpsi$ lies on the border of the acceptable region and corresponds to 
$\balpha=(7,7)^\prime$.
The first row shows a comparison between our Bayesian estimates $\bpsi_1^{PMC}$, 
and $\bpsi_1^{CML}$, the complete maximum likelihood estimates. 
These estimates are obtained using the true values of the
latent variables $z_i$, thus bringing the problem back to a multivariate 
normal regression model 
of $\by$ over $|\bz|$, in which $\bxi$ and $\bpsi$ play the roles of an intercept and a slope.
In fact, for known values of $\bz|$, the complete likelihood function reduces to
\begin{displaymath}
L(\btheta; \by, \bz )\propto |\bG|^{-n/2}\exp \left\{ -\frac 12 \sum_{i=1}^n 
 (\by_i-\bxi-\bpsi|z_i|)^\prime \bG^{-1}(\by_i-\bxi-\bpsi|z_i|) \right\},
\end{displaymath}
and the conditional (on $z$'s) maximum likelihood (CML) estimates are:
\begin{eqnarray*}
 \hat\bpsi^{CML}&=&\frac{\sum_i (\vert z_i \vert (\by_i - \hat \bxi^{CML}))}
{\sum_i z_i ^2 }; \\
 \rule{0cm}{0.5cm}\hat\bxi^{CML}&=&\bar{\bf y}-\hat\bpsi^{CML}\overline{|\bz|};\\
 \rule{0cm}{0.7cm}\hat \bG^{CML} &=& n^{-1} 
(\overline{\by}-\hat\bxi^{CML}-\hat\bpsi^{CML}\overline{|\bz|})
(\overline{\by}-\hat\bxi^{CML}-\hat\bpsi^{CML}\overline{|\bz|})'.
\end{eqnarray*}
The CML estimator is to be considered as a benchmark, as it uses an additional piece of
information which is not available for ML and PMC estimators. \\
The PMC estimates are concentrated in a single cloud around the true value and they
are in close agreement with the CML estimates. Very few points fall far from the cloud: it is probably a
consequence of the multimodality of the posterior distribution.

The second row shows the comparison
between ML estimates $\hat\bpsi_1^{ML}$ and $\hat\bpsi_1^{CML}$.
The dashed line is the bisector of the first and third quadrant, and the solid line represents
the true value of the parameter.
Scatterplots of the ML estimates reveal the odd behaviour of the likelihood, with points
in the ``genuine'' part of the distribution showing a higher variability.\\
The last row of Figure \ref{fig:Confronti} show the bivariate scatterplots of the Bayesian point estimates 
$\hat\bpsi^{PMC}$ and of the maximum likelihood estimates $\hat\bpsi^{ML}$, with lines indicating the true values of the parameters.
The sampling
distribution of $\hat\bpsi^{ML}$ is clearly multimodal. It also shows a larger dispersion and a negative skewness for both $\hat\psi_1^{ML}$ 
and $\hat\psi_2^{ML}$.

\begin{figure}[ht]
\centerline{\includegraphics[angle=-0,width=0.7\textwidth]{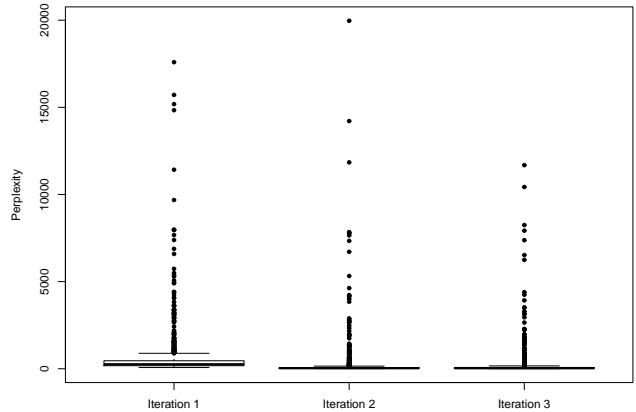}}
 \label{perpl}
 \caption{Perplexity measures in the first three iterations of the PMC algorithm for a simulation. 
It shows a typical pattern, with a decreasing number of outlier particles.}
\end{figure}

\subsection{A real dataset}
As a final illustration of the proposed algorithm,
we analyse the returns of two stocks in the NYSE composite index, namely the ``ABM
Industries Incorporated'' and ``The Boeing Company'' for the two decades October 1, 1992 to
October 1, 2012 (240 monthly observations). Data are available at \\
\verb-http://finance.yahoo.com/q/cp?s=%5ENYA+Components-.
Data show a moderate degree of skewness.

\begin{figure}[htb]
 \includegraphics[width=\textwidth]{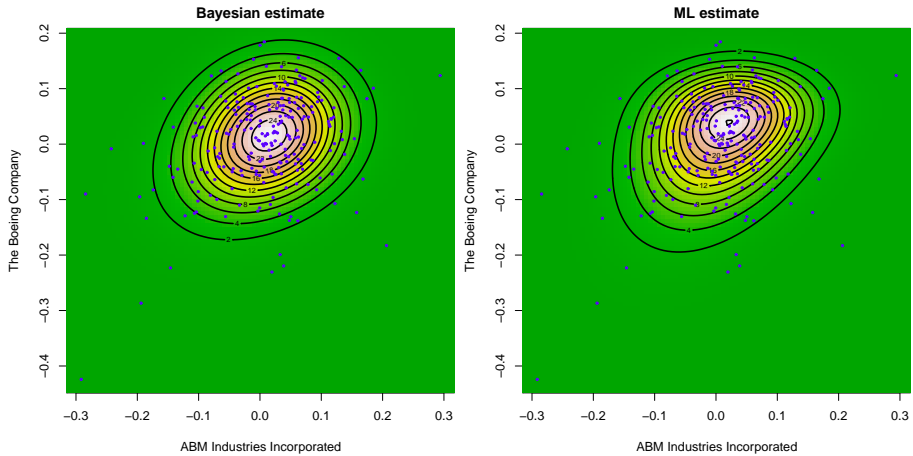}
 \caption{Returns of the stocks and contour plot of the density of the estimated $SN_2$ 
model, using the PMC algorithm (left) and the ML approach (right).}
 \label{estDens}
\end{figure}

We have performed a PMC sampler with $25$ iterations, $30,000$ particles each. 
Figure \ref{estDens} displays the raw data and the estimated $SN_2$ density obtained from
our proposed algorithm (left) and from the ML approach (right). 
Table \ref{estTab} summarizes the marginal posterior distributions of the parameters.% and Figure \ref{psi} displays the resampled particles of $(\psi_1, \psi_2)$ at the final iteration of the PMC algorithm.

\begin{table}[ht]
 \begin{center}
 \begin{tabular}{rrrrrrrr}
   \hline
   & $\xi_1$ & $\xi_2$ & $\rho$ & $\omega_1$ & $\omega_2$ & $\psi_1$ & $\psi_2$ \\ 
   \hline
   1\% & 0.040 & 0.042 & 0.310 & 0.085 & 0.086 & -0.075 & -0.076 \\ 
  5\%  & 0.045 & 0.049 & 0.369 & 0.088 & 0.090 & -0.072 & -0.076 \\ 
  50\% & 0.059 & 0.061 & 0.462 & 0.096 & 0.095 & -0.064 & -0.065 \\ 
  95\% & 0.068 & 0.068 & 0.526 & 0.101 & 0.102 & -0.046 & -0.054 \\ 
  99\% & 0.071 & 0.072 & 0.559 & 0.104 & 0.104 & -0.042 & -0.048 \\ 
   \hline
  \end{tabular}
 \end{center}
 \caption{Marginal posterior quantiles for the parameters of the model.}
 \label{estTab}
\end{table}

\begin{figure}[htb]
 \includegraphics[width=\textwidth]{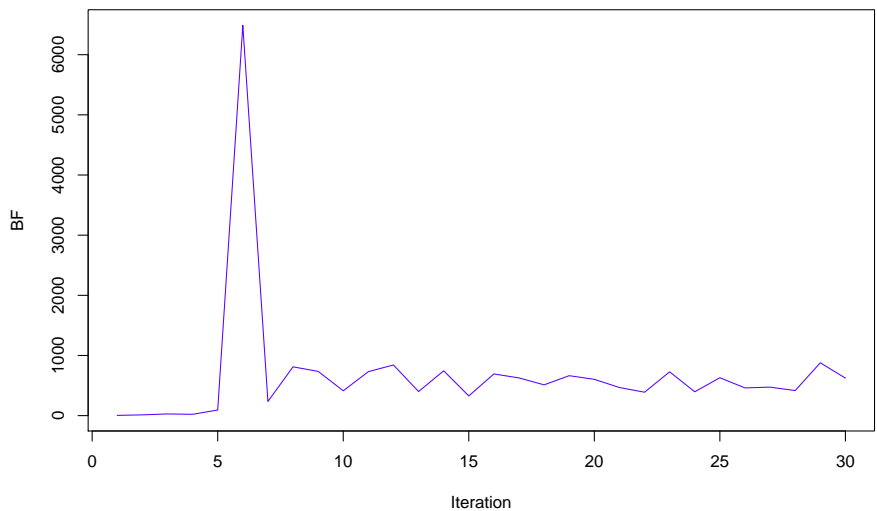}
 \caption{Pattern of the estimated Bayes factor over iterations.}
 \label{normcost}
\end{figure}

Figure \ref{normcost} depicts a typical pattern of an MC estimate of the Bayes factor
throughout the iterations: 
at the sixth
iteration, a huge jump occurs, probably due to the discovery of a region of
high posterior density. This causes a degeneracy of the particles, the production of non
reliable estimates and a rise in the perplexity index for that iteration. Once that the
new region has been explored, the estimates become stable. The high value of the estimated
Bayes factor in the
sixth iteration will not affect the final estimate, as it will be downweighted through the
perplexity index. Using formula (\ref{bbff}), the final estimate of the Bayes factor is
$\hat B_{10}=500.76$, showing overwhelming evidence in favour of the skew-normal model
compared to the normal one.

\section{Acknowledgements}
\label{sec:grants}
The Authors are sincerely grateful to an Associate Editor and two anonymous referees whose 
constructive criticism greatly improved a first version of the present paper.
Work supported by the project PRIN 2008: New developments in Bayesian sampling: theory and practice,
Project number 2008CEFF37, Sector: Economics and Statistics.

% Bibliography
%\section{Bibliography}
\bibliographystyle{elsarticle-harv}
\bibliography{skewbib}

\end{document}